\documentclass[twocolumn,superscriptaddress,nofootinbib,tightenlines,showpacs,showkeys,floatfix,preprintnumbers]{revtex4}

\usepackage[latin1]{inputenc}
\usepackage{epsfig,bm}
\usepackage{amsmath,amsfonts,amssymb,amscd,amsxtra,amsthm}
\usepackage{graphicx}
\usepackage[usenames]{color}
\usepackage[all]{xy}

\newcommand{\nn}{\nonumber}
\newcommand{\bra}{\langle}
\newcommand{\ket}{\rangle}

\begin{document}

\preprint{TUM-T39-08-08, MIT-CTP-3938}

%%%%%%%%%%%%%%%%%%%%% Title %%%%%%%%%%%%%%%%%%%%%%

\title{QCD sum rules for $\rho$ mesons in vacuum and in-medium, re-examined}

%%%%%%%%%%%%%%%%%%%% Authors %%%%%%%%%%%%%%%%%%%%%
\author{Youngshin Kwon}
\email{ykwon@ph.tum.de}

\affiliation{Physik-Department, Theoretische Physik, Technische
Universit\"at M\"unchen, \\D-85747 Garching, Germany}

\author{Massimiliano Procura}
\email{mprocura@mit.edu}

\affiliation{Center for Theoretical Physics, Laboratory for Nuclear
Science, Massachusetts Institute of Technology, \\Cambridge,
Massachusetts 02139, USA}

\author{Wolfram Weise}
\email{weise@ph.tum.de}

\affiliation{Physik-Department, Theoretische Physik, Technische
Universit\"at M\"unchen, \\D-85747 Garching, Germany}

%%%%%%%%%%%%%%%%%%%% Abstract %%%%%%%%%%%%%%%%%%%%%

\begin{abstract}
An updated investigation of QCD sum rules for the first two moments
of $\rho$ meson spectral functions, both in vacuum and in-medium, is
performed with emphasis on the role of the scale related to
spontaneous chiral symmetry breaking in QCD. It is demonstrated that
these lowest moments of vector current spectral distributions do
permit an accurate sum rule analysis with controlled input including
QCD condensates of the lowest dimensions, whereas higher moments are
subject to uncertainties from higher dimensional condensates.
Possible connections with Brown-Rho scaling are discussed. The
factorization approximation for four-quark condensates is shown not
to be applicable in any of the cases studied.
\end{abstract}


\maketitle

\section{Introduction}

As the lowest ``dipole'' excitations of the QCD vacuum, the light
vector mesons (the $\rho$ meson, in particular) have traditionally
played an important prototype role in calculations and discussions
based on QCD sum rules \cite{SVZ79}.  In-medium versions of these
sum rules have been used to set constraints on the way in
which vector meson masses undergo possible changes in dense and hot
hadronic matter \cite{HL92,HLS95,KKW97}. Questions were raised, however,
concerning the interpretation of such studies. In-medium changes of meson
properties, such as their mass shifts in nuclear matter, have their
primary origin in long-distance physics described by meson-nucleon forward
scattering amplitudes \cite{EI97PRL} and not in the short-distance
physics represented by subleading terms of the operator product expansion
(see also related discussions in Refs.\cite{HL97,EI97}). In-medium QCD sum rules
have nonetheless been further developed and applied over the years
\cite{LM98,DHP01,ZPK02,SL06}, including studies with emphasis on the
density dependence of four-quark condensates \cite{TZK05,K08}. The present
work aims in a different direction: namely, identifying the spontaneous chiral symmetry
breaking scale, $4\pi f_\pi \sim 1$ GeV, and its possible change with increasing
baryon density, in the context of QCD sum rules for the lowest moments
of the vector meson spectral functions.

The issue of in-medium changes of hadron properties persists as a
fundamental theme ever since the Brown-Rho (BR) scaling hypothesis
\cite{BR91} was launched, establishing a conceptual relationship
between the shifts of hadron masses in matter and the sliding scale
of spontaneous chiral symmetry breaking with changing thermodynamic
conditions. Investigations along these lines included various model calculations of
vector meson spectral functions at finite temperatures and baryon
densities (see Refs.\cite{RCW97,RW00,PM01} and further studies concerning BR
scaling in the context of in-medium QCD sum rules, e.g. in Ref.\cite{RRM06}).  Such calculations
were performed with the aim of understanding the ``low-mass
enhancements'' observed in dilepton spectra produced in high-energy
heavy-ion collisions by the CERES/NA45 \cite{Aga98} and NA60
\cite{NA60} experiments at the CERN SPS. These explorations,
primarily focused on the behaviour of the $\rho$ meson in the
strongly interacting hadronic medium, were conducted for a long time
with two seemingly opposing quests: whether there is an in-medium
shift of the $\rho$ meson; or on the other hand, whether the strong
collisional broadening of the spectral function due to interactions
of the $\rho$ meson with nucleons and mesons in the medium would
render the primary issue of a mass shift physically meaningless.

In the present paper we point out that playing the notions of ``mass
shift'' and ``broadening'' against one another may in fact not be
the proper question to ask. For resonant states such as the $\rho$
meson, which start out with a large decay width already in vacuum,
identifying a mass in an even broader in-medium spectral
distribution makes sense only in terms of the first moment of this
spectral distribution. For the two lowest spectral moments, however,
quite accurate statements can be made within the framework of QCD
sum rules, as we shall demonstrate. We propose therefore to abandon
the ``mass shift'' versus ``broadening'' dispute altogether and
concentrate on an analysis of spectral moments in the context of QCD
sum rules. Identifying the chiral symmetry breaking scale in such an
analysis, both in vacuum and in-medium, permits addressing and
examining the BR scaling hypothesis in a refined and better focused
way.

The strategy pursued in this paper is an update of previous
work \cite{KW99} which is in turn closely related to finite energy
sum rules (FESR) \cite{KPT83, MW00}. The advantage of these sum rules is
that they do not have to rely on the existence of a window of
stability for the Borel parameter usually employed in the sum rule analysis.
Caution must nevertheless be exercised with FESR's \cite{Ma98, BDPS06}
concerning their sensitivity to high-energy properties of spectral functions and
the detailed modeling of the transition between resonance and continuum regions,
a question that we shall also address. We concentrate here on the rho meson.
Starting with vacuum sum rules for the $\rho$ we recall how the
delineation of scales between resonance and continuum parts of the
spectral function can be related to the scale for spontaneous chiral
symmetry breaking, $4\pi f_\pi \simeq 1.2$ GeV (the ``chiral gap''),
where $f_\pi = 92.4$ MeV is the pion decay constant. In-medium sum
rules are examined using two complementary spectral functions as
generic examples: the one calculated in Ref.\cite{KKW97} using a
chiral meson-nucleon effective Lagrangian with vector mesons as
explicit degrees of freedom; and the one calculated in
Ref.\cite{RCW97} using a model which emphasizes the role of
particle-hole excitations including baryon resonances. Both types of
spectral functions were applied earlier \cite{RW00, RSW02} in
descriptions of the CERES/NA45 dilepton data \cite{Aga98}. Updated
versions of such spectral distributions have been used recently
\cite{RGR07, HR07} in comparisons with the more accurate NA60 data
\cite{NA60}.

\section{Reminder of QCD sum rules for isovector currents}

We begin with a brief introductory recollection of the QCD sum rule approach for excitations carrying the
quantum numbers of the $\rho$ meson, $J^{\pi} = 1^-$ and isospin $I=1$. The corresponding quark current  $j^\mu(x) = {1\over 2}(\bar{u}\gamma^\mu u -  \bar{d}\gamma^\mu d)$ figures in the current-current correlation tensor
\begin{equation}
  \Pi^{\mu\nu}(q)=i\int{d^4x}~e^{iq\cdot{x}}\bra\mathcal{T}j^\mu(x)j^\nu(0)\ket~~.
  \label{correlator}
\end{equation}
In vacuum this tensor can be reduced to a single scalar correlation function,
$\Pi(q^2) = {1\over 3} g_{\mu\nu}\Pi^{\mu\nu}$. In a nuclear medium the distinction needs to be made between longitudinal and transverse correlation functions. For vanishing three-momentum ($q^\mu = (\omega, \vec{q} = 0)$, the case considered here throughout), the longitudinal and transverse correlation functions coincide and will again be denoted as $\Pi(\omega,  \vec{q} = 0)$.

The next step is to write $\Pi(q^2)$ as a twice subtracted dispersion relation:
\begin{equation}
  \Pi(q^2)=\Pi(0)+\Pi'(0)\,q^2+{q^4\over\pi}\int{ds}{{\mathrm{Im}\Pi(s)}\over{s^2(s-q^2-i\epsilon)}}~~.
\label{dispersion}
\end{equation}
Alternatively, the same quantity is expressed at large spacelike $q^2 = -Q^2 < 0$ in terms of the Wilson operator product expansion (OPE):
\begin{equation}
  \begin{split}
    &12\pi^2\,\Pi(q^2=-Q^2)\\
    &\qquad=-c_0\,Q^2\ln\Big({{Q^2}\over{\mu^2}}\Big)+c_1+{{c_2}\over{Q^2}}+{{c_3}\over{Q^4}}+\,\cdots~~.
 \label{ope}
  \end{split}
\end{equation}
In vacuum and for the $\rho$ meson channel, the expansion coefficients are given as:
\begin{equation}
  \begin{split}
    c_0&={3\over2}\Big(1+{{\alpha_s}\over{\pi}}\Big) + \,\,\cdots~~,\\
    c_1&= -{9\over2}(m^2_u+m^2_d)~~,\\
    c_2&={{\pi^2}\over2}\Big\bra{{\alpha_s}\over\pi}G^2\Big\ket+6\pi^2\big(m_u\bra\bar{u}u\ket+m_d\bra\bar{d}d\ket\big)~~.
    \label{c_n}
  \end{split}
\end{equation}
These three leading coefficients are well determined. The dominant perturbative QCD piece $c_0$ is shown here including just the standard ${\cal O}(\alpha_s)$ correction. At a later stage and in all explicit calculations, the QCD corrections will be further extended up to and including ${\cal O}(\alpha_s^3)$
(see Appendix A).

The quark mass term $c_1$ is small and can safely be neglected. The coefficient $c_2$ involves the QCD condensates of lowest dimension four.  The quark condensate times the quark mass is given accurately through the Gell-Mann - Oakes - Renner relation as
\begin{equation}
 \begin{split}
\bra m_u\,\bar{u}u +m_d\,\bar{d}d\ket &\simeq m_q\bra \bar{u}u+\bar{d}d\ket\\
&= -m_\pi^2\,f_\pi^2 = - (0.11\, \text{GeV})^4~~.
 \end{split}
 \label{GOR}
\end{equation}
The gluon condensate $\bra(\alpha_s/\pi)\,G^2\ket \sim (0.3$
GeV$)^4$ is (far less accurately)  determined by charmonium sum
rules. For a detailed discussion see Ref.\cite{Io06} where an upper
limit
\begin{equation}
\bra(\alpha_s/\pi)\,G^2\ket^{1/4} \lesssim 0.31 \, \text{GeV}
\nonumber
\end{equation}
is given.

In-medium corrections to leading order in the baryon density $\rho$ are introduced by the replacement $c_2\rightarrow c_2 + \delta c_2(\rho)$, with \cite{HL92,HLS95,KKW97}
\begin{equation}
  \delta{c_2}=3\pi^2\big[A_1M_N-{4\over27}M^{(0)}_N+2\sigma_N\big]\,\rho~~.
  \label{c_2med}
\end{equation}
The first term in brackets is the leading density dependent
perturbative QCD correction. It involves the first moment, $A_1 =
2\langle x \rangle_{u+d}$, of the parton distribution in the
nucleon. Given the empirical (MRST) \cite{MRS98,MRS01} momentum
fraction carried by $u$ and $d$ quarks in the nucleon, $\langle x
\rangle_{u+d} \simeq 0.62$ at $Q^2 = 1$ GeV$^2$, we use $A_1 \simeq
1.24$ (see Appendix B).

The second term on the r.h.s. of Eq.(\ref{c_2med}) is the correction
to the gluon condensate  at finite density. It is proportional to
the nucleon mass in the chiral limit for which we use  $M_N^{(0)} =
0.88$ GeV from Ref.\cite{PMW06}. The third term represents the
leading density dependence of the quark condensate. It is
proportional to the nucleon sigma term, $\sigma_N = (45\pm 8)$ MeV
\cite{GLS}. By far the largest contribution to $\delta c_2$
evidently comes from the $A_1$ term, so that the large uncertainty
in $\sigma_N$ has only relatively minor consequences.

Following these considerations the input for $c_2$ and $\delta c_2$ is summarized in Table I. The in-medium sum rule analysis will be done at normal nuclear matter density, $\rho = \rho_0 = 0.17$ fm$^{-3}$.

\begin{table}[h]
 \begin{tabular}{c|c|c}
  \hline
       &   value   &   reference\\
  \hline
   $M_N$    &    $939\,\mathrm{MeV}$    & \\
   $m_q\bra\bar{q}q\ket$    &    $-(0.11\,\mathrm{GeV})^4$    &    GOR\\
   $\bra{{\alpha_s}\over{\pi}}G^2\ket$    &    $\,\,0.005\pm0.004\,\mathrm{GeV}^4\,$    &    \cite{Io06}\\
   $A_1$    &    $1.237$ & \cite{MRS01}\\
   $M^{(0)}_N$    &    $0.88\,\mathrm{GeV}$    &    \cite{PMW06}\\
   $\sigma_N$    &    $45\pm 8\,\mathrm{MeV}$    &  \cite{GLS}\\
 \hline
 \end{tabular}
 \label{ta:input}
 \caption{Input summary}
\end{table}

The coefficient $c_3$ involves four-quark condensates in the following combination:
\begin{equation}
  \begin{split}
    c_3&=-6\pi^3\alpha_s\big[\bra(\bar{u}\gamma_\mu\gamma_5\lambda^au-\bar{d}\gamma_\mu\gamma_5\lambda^ad)^2\ket\\
    &\quad+{2\over9}\bra(\bar{u}\gamma_\mu\lambda^au+\bar{d}\gamma_\mu\lambda^ad)\sum_{q=u,d,s}\bar{q}\gamma^\mu\lambda^aq\ket\big]
    \label{c_3}
  \end{split}
\end{equation}
These condensates of dimension six are not known at any reasonable level of precision. What is commonly done at this point is to introduce a factorization approximation, truncating intermediate
states by the QCD ground state and writing
\begin{equation}
    \bra(\bar{q}\gamma_\mu\gamma_5\lambda^a{q})^2\ket=-\bra(\bar{q}\gamma_\mu\lambda^a{q})^2\ket
    ={16\over9}\kappa\,\bra\bar{q}q\ket^2~~,
\label{4q}
\end{equation}
with $\kappa$ introduced to parametrize deviations from exact
factorization ($\kappa = 1$). The in-medium analogue including terms
linear in the density $\rho$ becomes
\begin{equation}
  c_3=-{448\over27}\kappa(\rho)\,\pi^3\alpha_s\Big(\bra\bar{q}q\ket^2+{{\sigma_N\,\bra\bar{q}q\ket}\over{m_q}}\rho\Big)~~,
  \label{c_3fac}
\end{equation}
with a density dependent $\kappa$ parameter.

Clearly, a QCD sum rule analysis that aims for accuracy must try to avoid the uncertain four-quark condensate piece $c_3$ in the OPE hierarchy. This is indeed possible when considering only the two lowest moments of the spectral function, Im$\Pi(s)$, as follows. We introduce the dimensionless spectral function
\begin{equation}
R(s) = -{12\pi\over s} \text{Im}\,\Pi(s)~~.
\end{equation}
Note that, in vacuum, $R(s)$ is identified with the observable $\sigma(e^+e^-\rightarrow \text{hadrons})/\sigma(e^+e^-\rightarrow\mu^+\mu^-)$. Now assume as usual that there exists a delineation scale $s_0$ which separates the low-mass resonance region ($s \leq s_0$) from the high-mass continuum ($s > s_0$):
\begin{equation}
    R(s)=R_\rho(s)\,\Theta(s_0-s) +R_c(s)\,\Theta(s-s_0)~.
    \label{R_rho}
\end{equation}
This step function delineation between resonance and continuum seems schematic on first sight. In practice, the transition to the continuum is smooth and $s_0$ should be considered as an average
scale characterizing the transition region. A detailed analysis, to be described later, shows that the step function ansatz is equivalently as valid as a more realistic modeling of the threshold "ramp", e.g. by the dotted line in Fig.\ref{figure1}.

Let the high-mass continuum be subject to a perturbative QCD
treatment, following duality considerations:
\begin{equation}
R_c(s) \rightarrow c_0\,\,\,\,\,\text{for}\,\,\,\,\, s>s_0~~.
\end{equation}
Then perform a Borel transformation on Eqs.(\ref{dispersion}-\ref{ope}), leading to
\begin{equation}
  \begin{split}
    &12\pi^2\Pi(0)+\int^\infty_0{ds}\,R(s)\,e^{-s/\mathcal{M}^2}\\
    &\qquad=c_0\mathcal{M}^2+c_1+{{c_2}\over{\mathcal{M}^2}}+{{c_3}\over{2\mathcal{M}^4}}+\cdots
    \label{borelsr}
  \end{split}
\end{equation}
Choose the (otherwise arbitrary) Borel scale parameter sufficiently
large, ${\cal M}>\sqrt{s_0}$, expand $e^{-s/{\cal M}^2}$ and arrange
term by term in inverse powers of ${\cal M}$. The result is a
hierarchy of sum rules for $moments$ of the low-mass part of the
spectral function $R(s)$:

{\setlength\arraycolsep{2pt}
\begin{eqnarray}
  \int^{s_0}_0ds\,R_\rho(s)&=&s_0\,c_0+c_1-12\pi^2\,\Pi(0)~,
  \label{0momsr}\\
  \int^{s_0}_0ds\,sR_\rho(s)&=&{{s^2_0}\over2}c_0-c_2~,
  \label{1momsr}\\
  \int^{s_0}_0ds\,s^2R_\rho(s)&=&{{s^3_0}\over3}c_0+c_3~.
  \label{2momsr}
\end{eqnarray}
}These equations are written again to first order in $\alpha_s$,
with $c_0 = (3/2)(1+\alpha_s/\pi)$. Corrections to order
$\alpha_s^3$ are included by the replacements $c_0 \rightarrow c_0 +
(3/2)\varepsilon_n$ in the $n$-th moment, with $\varepsilon_n$ given
explicitly in Appendix A. In the detailed calculations the relevant
running coupling is to be taken as $\alpha_s(s_0)$ with $s_0\sim 1$
GeV$^2$, the onset scale for the (multipion) continuum part of the
quark-antiquark excitation spectrum. We use
\begin{equation}
\alpha_s(s_0\sim 1\, \text{GeV}^2) = 0.50\pm 0.03~,
\label{alpha}
\end{equation}
referring to the most recent NNLO $(\,\overline{\mathrm{MS}}\,)$
analysis in \cite{Be07,PRS07}. The error in $\alpha_s(s_0)$ is
actually the major source of uncertainty in the sum rule
calculation, all other corrections being considerably smaller in
magnitude relative to the leading term.

The subtraction constant $\Pi(0)$ in Eq.(\ref{0momsr}) vanishes in
vacuum. At finite density this is the Landau term, $\Pi(0) =
{\rho\over4M_N}$, analogous to the Thomson limit in Compton
scattering.

Note that even for a broad spectral distribution $R(s)$, a squared
``mass'' associated with the low-energy sector of this spectrum can
be defined through the ratio of the first and zeroth moments,
Eqs.(\ref{0momsr},\ref{1momsr}) (see also Ref.\cite{ZPK02}):
\begin{equation}
  \bar{m}^2={{\int^{s_0}_0{ds}\,sR(s)}\over{\int^{s_0}_0{ds}\,R(s)}}~~.
\end{equation}

\section{Vacuum sum rules}

\subsection{Identifying the spontaneous chiral symmetry breaking scale}

Consider now first the sum rule for the isovector current-current correlation function in vacuum.
Following Ref.\cite{MW00} we start from the working hypothesis that the scale $s_0$ delineating
low-energy and continuum parts of the vector-isovector quark-antiquark spectrum should be identified
with the scale for spontaneous chiral symmetry breaking in QCD:
\begin{equation}
\sqrt{s_0} = 4\pi f_\pi~~.
\label{chiscale}
\end{equation}
For illustration, recall the schematic (large $N_c$) example of a zero-width $\rho$ meson,
\begin{equation}
R_\rho(s)= {12\pi^2\,m_\rho^2\over g^2}\,\delta(s-m_\rho^2)~,
\end{equation}
with the vector coupling constant $g$. Neglecting small quark masses as well as QCD
and condensate corrections in Eqs.(\ref{0momsr},\ref{1momsr}), one arrives at
\begin{equation}
\begin{split}
 \int^{s_0}_0ds\,R_\rho(s)&={3\over 2}s_0\, = \,24\pi^2f_\pi^2~,\\
 \int^{s_0}_0ds\,sR_\rho(s)&={3\over 4}s^2_0\,=\,192\pi^4 f_\pi^4~,
\end{split}
\end{equation}
and immediately recovers a celebrated current algebra result
(the KSRF relation \cite{KSRF}),
\begin{equation}
m_\rho = \sqrt{2}\,g f_\pi~,
\end{equation}
together with the universal vector coupling $g = 2\pi$.

While this schematic example underlines the validity of the hypothesis (\ref{chiscale}), a more detailed
test using a realistic spectral distribution $R(s)$ and the full sum rule analysis, including corrections,
must of course be performed.  We do this along the lines of Ref.\cite{MW00} and update the results
found in that work.

\begin{figure}[h]
 \includegraphics[width=8.5cm]{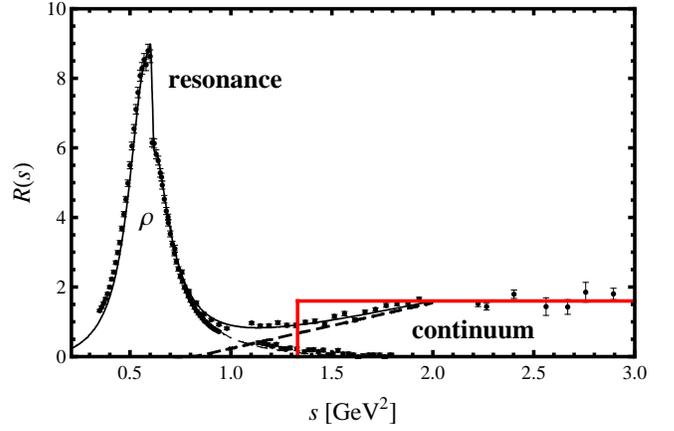}
 \caption{Vector-isovector spectral function in vacuum showing the $\rho$ resonance and continuum parts as described in the text and compared to $e^+e^- \rightarrow \pi^+\,\pi^-$ ($\rho$ resonance region) and  $e^+e^- \rightarrow n\,\pi$ data with $n$ even \cite{LNF05,Dol91}.}
\label{figure1}
\end{figure}

The input is now the resonant $\rho$ meson spectral function $R_\rho(s)$ calculated from
one-loop chiral $\pi\pi$ dynamics with gauge coupling to vector mesons \cite{KKW96,KKW97}.
The $n$-pion continuum $R_c(s)$ (with $n\geq 4$ even) is parametrized as in Eq.(\ref{R_rho}),
with the gap scale $s_0$ to be determined by the sum rules for the lowest two moments, Eqs.(\ref{0momsr},\ref{1momsr}). The spectral function $R(s)$ is shown in comparison with experimental data
in Fig.\ref{figure1}.

The analysis proceeds as follows. The equations for the two lowest
moments of $R(s)$, {\setlength\arraycolsep{2pt}
\begin{eqnarray}
  \int^{s_0}_0ds\,R_\rho(s)&=&s_0\,\left(c_0+ {3\over 2}\varepsilon_0\right) + c_1~,
  \label{0mom}\\
  \int^{s_0}_0ds\,sR_\rho(s)&=&{{s^2_0}\over2}\left(c_0+ {3\over 2}\varepsilon_1\right)-c_2~,
  \label{1mom}
\end{eqnarray}
are solved to determine $s_0$. For the zeroth moment Eq.(\ref{0mom}) gives
$\sqrt{s_0} = 1.13\pm0.02$ GeV. Overall consistency requires that
the same $s_0$ results also from Eq.(\ref{1mom}) within an error band
determined by the uncertainties of the input summarized in table I
and Eq.(\ref{alpha}). This test turns out to be successful. The
detailed analysis of uncertainties performed with
Eq.(\ref{1mom}) for the first moment is shown in
Fig.\ref{figure2}. The resulting $\sqrt{s_0} = 1.14\pm0.01$ GeV is
within 2\% of the empirical $4\pi f_\pi \simeq 1.16$ GeV using the
physical value $f_\pi = 92.4$ MeV of the pion decay constant. The
postulate (\ref{chiscale}) identifying $\sqrt{s_0}$ with the scale
characteristic of spontaneously broken chiral symmetry, appears to
be working quantitatively.

\begin{figure}[h]
 \includegraphics[width=8.5cm]{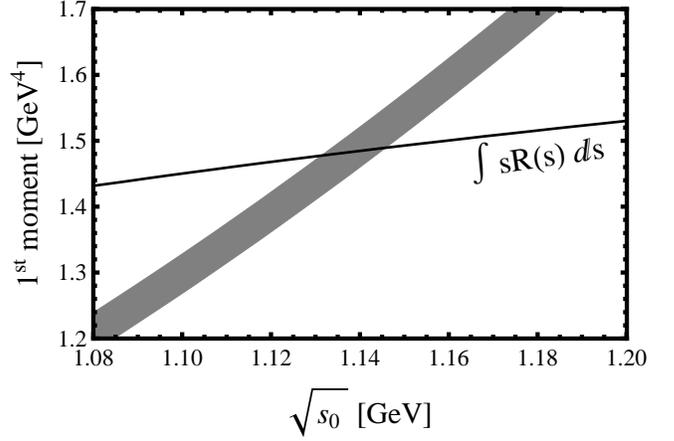}
 \caption{QCD sum rule analysis of the $\rho$ meson spectral function in vacuum. First moment (solid line, left-hand side of Eq.(\ref{1mom})) is plotted versus right-hand side
 (grey band including uncertainties) as function of the gap scale $\sqrt{s_0}$ delineating
 low-mass resonance region from high-mass continuum.}
\label{figure2}
\end{figure}

The relation between first and the zeroth moment,
\begin{equation}
\int^{s_0}_0ds\,s R_\rho(s) = {\cal F}(s_0)\,\int^{s_0}_0ds\,R_\rho(s)
\label{ratio}
\end{equation}
thus involves a uniquely determined function of $s_0$:
\begin{equation}
{\cal F}(s_0) = {s_0^2\left(c_0+{3\over
2}\varepsilon_1\right)-2c_2\over 2s_0\left(c_0+{3\over
2}\varepsilon_0\right)+2c_1}~~, \label{func}
\end{equation}
up to the estimated uncertainties in the quantities $c_i$ and
$\varepsilon_n$ (the largest error being  associated with
$\alpha_s(s_0)$). The squared mass given by $\bar{m}_\rho^2 = {\cal
F}(s_0) \simeq 0.611\pm0.013$ GeV$^2$ or $\bar{m}_\rho \simeq
0.78\pm0.01$ GeV, is very close to the physical $\rho$ meson mass as
expected. In fact the canonical relation $\bar{m}_\rho =
\sqrt{s_0/2} =\sqrt{2}\cdot 2\pi f_\pi$ turns out to be satisfied
again at the 2\% level, demonstrating the smallness of the
next-to-leading QCD corrections and of the condensate term $c_2$.

\subsection{Sensitivity to continuum threshold modeling}
\label{thrmod}

The question arises whether the quantitatively successful identification of the
continuum threshold $\sqrt{s_0}$ with the chiral symmetry breaking scale (i.e.
the consistency of the QCD sum rule analysis with current algebra results)
is influenced by the schematic step-function
parametrization (\ref{R_rho}). A test can be performed replacing the step function by a ramp
function to yield a smooth transition between resonance and continuum region, as follows:
\begin{equation}
     R(s)=R_\rho(s)\,\Theta(s_2-s)+R_c(s)\,W(s)~,
     \end{equation}
where the weight function, $W(s)$, is defined as
\begin{equation}
  W(x)=\left\{\begin{array}{cl}
           \vspace{2mm}
          0 & \text{ for } x\leq s_1\\
           \vspace{1.5mm}
          \displaystyle{{x-s_1}\over{s_2-s_1}} & \text{ for }s_1\leq x\leq s_2\\
          1 & \text{ for } x\geq s_2~.
        \end{array}\right.
  \label{eq:Wofs}
\end{equation}
The step function behavior is recovered for $W(x)$ in the limit $s_1\rightarrow s_2$.

Using the function $W(s)$, the modified sum rules for the lowest two moments
of the spectrum $R(s)$ become
\setlength\arraycolsep{2pt}
\begin{eqnarray}
   \int^{s_2}_0ds\,R_{\rho}(s)&=&s_2\left(c_0+{3\over2}\varepsilon_0\right)+c_1-12\pi^2\Pi(0)\nn\\
                            &&-\left(c_0-R_{\rho}(s_2)\right)\int^{s_2}_{s_1}ds\,W(s)~,
              \label{0thsr}\\
   \int^{s_2}_0ds\,sR_{\rho}(s)&=&{{s^2_2}\over{2}}\left(c_0+{3\over2}\varepsilon_1\right)-c_2\nn\\
                             &&-\left(c_0-R_{\rho}(s_2)\right)\int^{s_2}_{s_1}ds\,sW(s)~.\quad
              \label{1stsr}
\end{eqnarray}
Sets of intervals $[s_1, s_2]$ are then determined so as to satisfy both sum rules (\ref{0thsr},\ref{1stsr}), and the scale $s_0$ defined by
\begin{equation}
   s_0={{s_1+s_2}\over{2}}~,
   \label{eq:s0}
\end{equation}
is now introduced to characterize the continuum threshold.  As shown in Fig.\ref{figure6},
the resulting $\sqrt{s_0}$ is stable with
respect to variations in the slope $(s_2-s_1)^{-1}$ of the ramp function
$W(s)$, thus confirming that the step function parametrization of the continuum
is not restrictive: the smooth ``ramping" into the continuum\footnote{In this test the uncertainties of $\alpha_s(Q^2)$ and of the gluon condensate have been excluded for simplicity.}
produces values of $\sqrt{s_0}$ that fall within the narrow (less than 1 \%) uncertainty band of the step function approach. We note at this point that the best fit to the empirical spectral function
has  $s_2-s_1\simeq1~\mathrm{GeV}^2$ (see Fig.\ref{figure1}). It can be concluded that the present sum rule analysis and the observed quantitative agreement of the continuum threshold with the chiral gap $4\pi f_\pi$ do not depend on details of the threshold modeling.}
\begin{figure}[ht]
   \includegraphics[width=8.5cm]{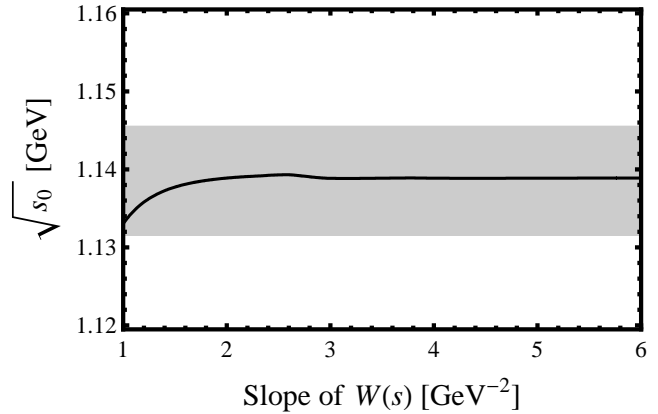}
   \caption{Dependence of $\sqrt{\,s_0}$ (determined from Eqs.(\ref{0thsr}-\ref{eq:s0})) on the slope
    $(s_2-s_1)^{-1}$ of  the ramp function $W(s)$ describing the onset of the continuum in the vacuum sum rule. The grey band indicates the uncertainty range of the result obtained with step function parametrization of the continuum.}
   \label{figure6}
\end{figure}

\section{In-medium sum rules}

In this section the approach just described is applied analogously
to vector current spectral functions at finite density. We start
again from Eqs.(\ref{0mom},\ref{1mom}), now with inclusion of
$\Pi(0) = {\rho\over4M_N}$ and the density dependent corrections to
the condensate terms, $c_2\rightarrow c_2 + \delta c_2$ (see
Eq.(\ref{c_2med})).

\begin{figure}[h]
 \includegraphics[width=8.5cm]{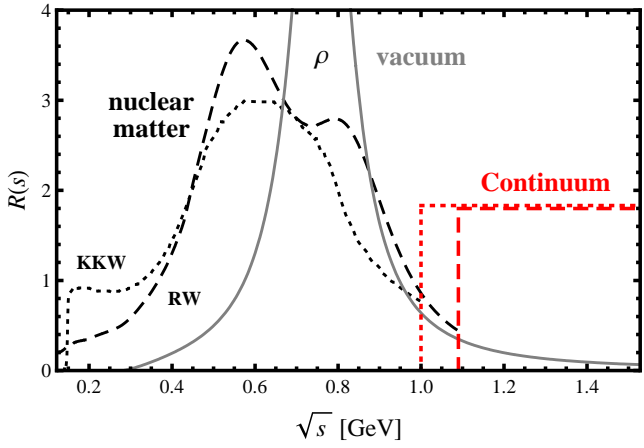}
 \caption{In-medium isovector vector spectral functions at nuclear matter density,
 $\rho_0 = 0.16$ fm$^{-3}$, taken from Refs.\cite{KKW97} (KKW) and \cite{RW00} (RW). The $\rho$ meson spectrum in vacuum is also shown for comparison.}
\label{figure3}
\end{figure}

Two generic prototypes of in-medium isovector vector spectral
functions, Im$\Pi(\omega = \sqrt{s}, \vec{q} = 0; \rho)$, are used
for demonstration: the one derived from a chiral effective
Lagrangian with vector meson couplings constrained by vector
dominance \cite{KKW97} (referred to as KKW), and the one calculated
with emphasis on particle-hole excitations incorporating baryon
resonances \cite{RW00} (referred to as RW). The analysis is
performed at the baryon density of normal nuclear matter, $\rho =
\rho_0 = 0.17$ fm$^{-3}$. The KKW and RW spectral functions, taken
at this density, are shown in comparison in Fig.\ref{figure3}.

The KKW and RW in-medium spectral distributions both consistently
show a strong broadening as compared to the vacuum $\rho$ meson.
They differ in details at the low mass end of the spectrum. While
KKW emphasizes the role of chiral in-medium $\pi\pi$ interactions,
RW focuses on the role of nucleon-hole, $\Delta(1232)$-hole and
$N^*(1520)$-hole excitations. At first sight, none of these
broad distributions permit identifying an ``in-medium mass'' or a
shift thereof with respect to the $\rho$ meson mass in vacuum. This
has generally led to the conclusion of there being no $\rho$ mass
shift at finite density, but just an overwhelmingly large inelastic
width due to interactions of the coupled $\rho \leftrightarrow
\pi\pi$ system with nucleons in the nuclear medium.

We now perform the sum rule analysis, first with step function continuum, for the two leading moments of
the KKW and RW spectral distributions: {\setlength\arraycolsep{2pt}
\begin{eqnarray}
  \int^{s_0^*}_0 ds\,R_\rho(s)&=&s_0^*\,\left(c_0+ {3\over 2}\varepsilon_0\right) + c_1-{3\pi^2\rho\over
  M_N}~,
  \label{0mommed}\\
  \int^{s_0^*}_0 ds\,sR_\rho(s)&=&{{{s_0^*}^2}\over2}\left(c_0+ {3\over 2}\varepsilon_1\right)-\left(c_2 + \delta
  c_2(\rho)\right)\quad
  \label{1mommed}
\end{eqnarray}
where the gap scale $\sqrt{s_0^*}$ is permitted to adjust itself to
the in-medium situation. Consistency of the first and zeroth
spectral moments is again tested and observed to be satisfied within the
uncertainties of the input.  This determines $s_0^*$ at given density $\rho =
\rho_0$. Effects of smooth ramping into the continuum will again be examined later.

Fig.\ref{figure4} shows the outcome of this procedure for the KKW
spectral function. In this case, at nuclear matter density $\rho_0$,
the in-medium gap scale $\sqrt{s_0^*}$ is indeed seen to be shifted
downward from its vacuum position, $\sqrt{s_0} \simeq 1.14$ GeV
$\simeq 4\pi f_\pi$. One finds
\begin{equation}
\sqrt{s_0^*} = (1.00 \pm 0.02)\,\text{GeV}~~~~\text{(KKW at } \rho =
\rho_0)~.
\label{KKW}
\end{equation}
For comparison, the cross check with the sum rule for the zeroth moment gives
$\sqrt{s_0^*} = (1.02 \pm 0.03)$ GeV, consistent with (\ref{KKW}).

\begin{figure}[h]
 \includegraphics[width=8.5cm]{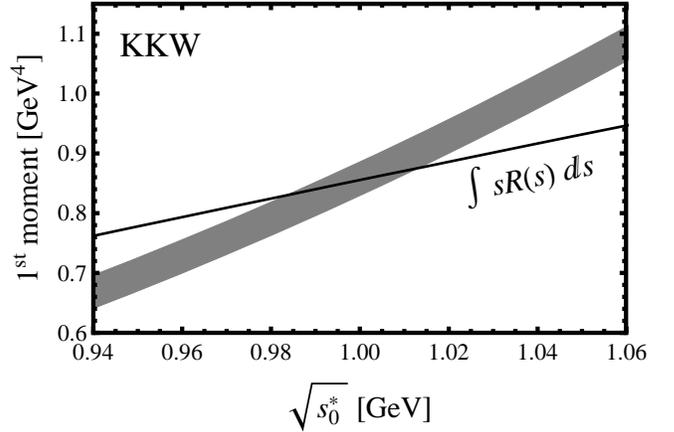}
 \caption{QCD sum rule analysis of the KKW in-medium spectral function
 \cite{KKW97}. First moment (solid line, left-hand side of Eq.(\ref{1mommed})) is plotted versus right-hand side (grey band including uncertainties) as function of the in-medium gap scale $\sqrt{s_0^*}$.}
\label{figure4}
\end{figure}
The analogue of Eq.(\ref{ratio}) becomes:
\begin{equation}
\int^{s_0^*}_0ds\,s R(s,\rho) = {\cal F}(s_0^*,\rho)\,\int^{s_0^*}_0ds\,R(s,\rho)
\label{ratiomed}
\end{equation}
with
\begin{equation}
{\cal F}(s_0^*,\rho) = {{s_0^*}^2\left(c_0+{3\over
2}\varepsilon_1\right)-2(c_2+\delta c_2(\rho))\over
2\left[s_0^*\left(c_0+{3\over
2}\varepsilon_0\right)+c_1-3\pi^2\rho/M_N\right]}~~, \label{funcmed}
\end{equation}
The average in-medium ``mass'' determined from the ratio
${\cal F}(s_0^*,\rho)$ of the first and zeroth spectral moments
is found to be
\begin{equation}
\bar{m}^*(\rho) = \sqrt{{\cal F}(s_0^*,\rho)} = (0.67\pm
0.02)\,\text{GeV}
\end{equation}
for the KKW spectral function at density $\rho = \rho_0$. One notes now that the ratio of in-medium and vacuum 1st spectral moments
behaves as
\begin{equation}
{\bar{m}^*\over \bar{m}_\rho(vac)} = \sqrt{{\cal
F}(s_0^*,\rho)\over{\cal F}(s_0,\rho=0)}\simeq 0.85\pm0.02
\end{equation}
at $\rho = \rho_0$.

The successful identification $\sqrt{s_0} = 4\pi f_\pi$ in
vacuum suggests a corresponding generalization to the
in-medium case: $\sqrt{s_0^*} = 4\pi f_\pi^*$, in terms of the pion
decay constant, $f_\pi^*\equiv f_t(\rho)$, related to the time component of the axial current at
finite density. Then one observes $\sqrt{s_0^*/ s_0} = f_\pi^*/f_\pi
= 0.88\pm 0.02$.  One finds, within uncertainties,
\begin{equation}
{\bar{m}^*\over \bar{m}_\rho(vac)} \simeq {f_\pi^*\over f_\pi}\sim 1 - (0.15\pm 0.02){\rho\over\rho_0}~,
\end{equation}
suggesting that the BR scaling tendency is indeed visible for the KKW in-medium spectral function, contrary to first impression when looking just at the very broad overall spectral distribution \cite{KKW97}.
In this context we refer to the subsequent section for an update on the relationship between the in-medium pion decay constant and the density dependence of the chiral condensate.

The KKW spectrum is based entirely on chiral pion dynamics with
vector mesons. Baryon resonances are assumed to develop large widths
and ``dissolve'' in nuclear matter so that they become part of the
continuous background. In contrast, the RW spectral function starts
from a different scenario in which baryon resonances play a
distinguished role, assuming that they maintain their quasiparticle
properties in matter. It is thus instructive to conduct, as before,  a corresponding
sum rule analysis for the moments of the RW spectrum under such
aspects.

The result is displayed in Fig.\ref{figure5}. One deduces
\begin{equation}
\sqrt{s_0^*} = (1.09 \pm 0.01)\,\text{GeV}~~~~\text{(RW at } \rho
= \rho_0)~
\label{RW}
\end{equation}
and $\sqrt{s_0^*/s_0}= 0.97\pm0.01$, together with ${\bar{m}^*\over
\bar{m}_\rho(vac)} \simeq 0.96\pm0.02$ at $\rho = \rho_0$. (For comparison,
the sum rule for the zeroth moment gives $\sqrt{s_0^*} = (1.11 \pm 0.02)$ GeV,
consistent with (\ref{RW})). So the RW
spectral function exhibits dominantly broadening with almost no
in-medium shift of the ratio of the moments. Notably, both RW and
KKW based spectral functions work quite well in comparison with
dilepton data taken at SPS energies (assuming models for the
expansion dynamics of the hot and dense matter which have their own
uncertainties). This implies that it is presumably not possible to
distinguish between the BR scaling scenario and other (opposing)
views from those data.

\begin{figure}[h]
 \includegraphics[width=8.5cm]{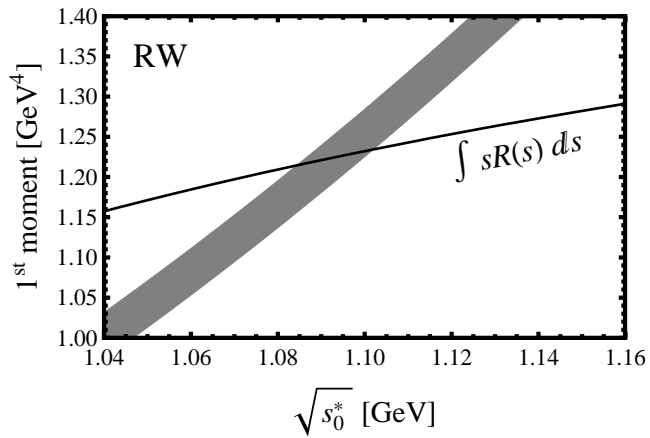}
 \caption{QCD sum rule analysis of the RW in-medium spectral function
 \cite{RW00}. First moment (solid line, left-hand side of Eq.(\ref{1mommed})) is plotted versus right-hand side (grey band including uncertainties) as function of the in-medium gap scale $\sqrt{s_0^*}$.}
\label{figure5}
\end{figure}

The ``ramping" test in order to establish stability with respect to
the modeling of the continuum is performed as for the vacuum case
described in the previous section, with the same ramping function
$W(s)$ employed also for the in-medium case. The results of this
test for the KKW and RW spectral functions are shown in
Fig.\ref{figure7}. One finds again that the determination of
$\sqrt{s_0^*}$, using a variety of smooth transitions to the
continuum, is insensitive to details of the threshold modeling
within the narrow band of uncertainties.
\begin{figure}[ht]
   \includegraphics[width=8.5cm]{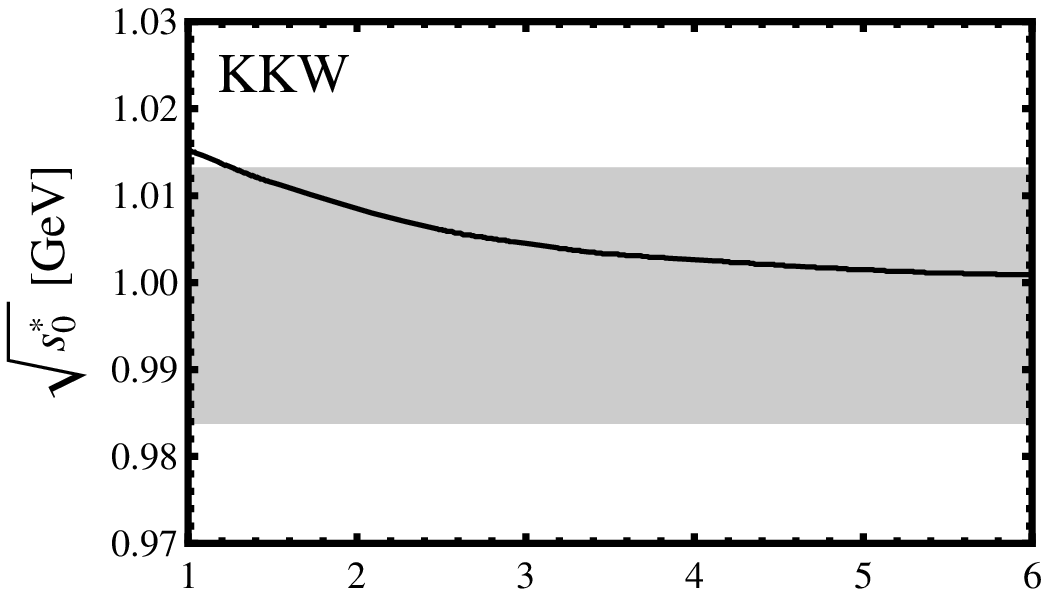}
   \includegraphics[width=8.5cm]{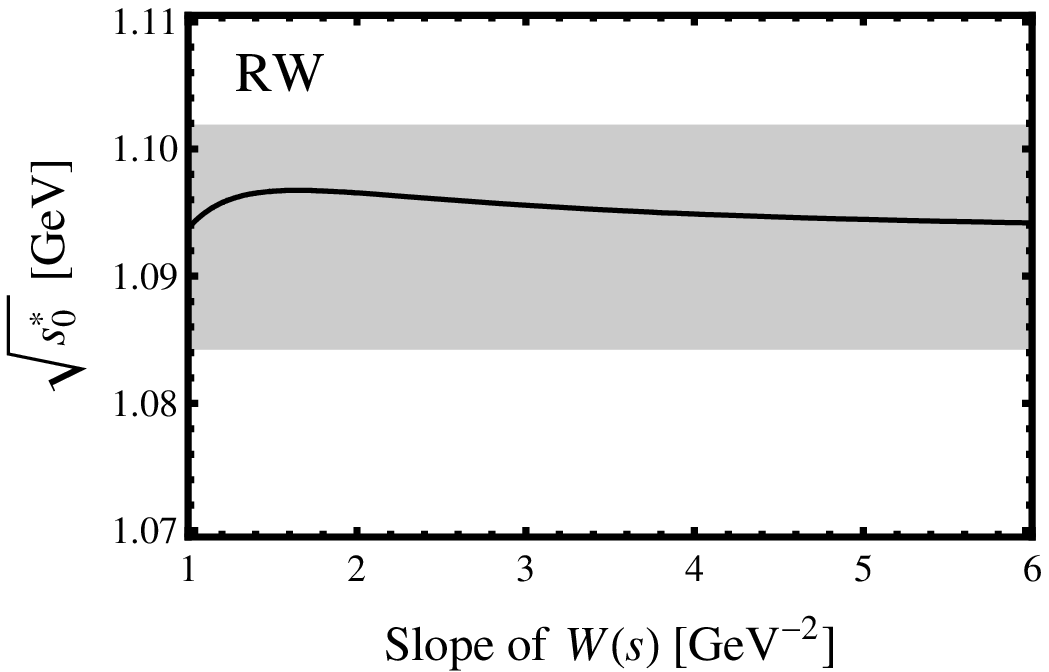}
   \caption{Dependence of $\sqrt{\,s_0^*}$, as in Fig.\ref{figure6}, on the slope
    $(s_2-s_1)^{-1}$ of  the ramp function $W(s)$, now describing the onset of the continuum in the in-medium sum rules. Upper panel: result for the KKW spectral function. Lower panel: for the RW spectral function. The grey bands indicate the uncertainty ranges of the results obtained with step function parametrizations of the continuum.}
   \label{figure7}
\end{figure}

\section{In-medium pion decay constant and chiral condensate: short digression}

The present QCD sum rule study asserts that the delineation
between low-energy resonance and high-energy continuum parts of the spectral function is related to the chiral scale, $4\pi f_\pi$, which acts as an order parameter for the spontaneously broken chiral symmetry of the QCD vacuum. Its in-medium change with increasing baryon density is of fundamental interest and deserves an added short section with an updated discussion.

In the nuclear medium, the relevant quantity is the pion decay constant $f_t(\rho) \equiv f_\pi^*(\rho)$ related to the time component of the axial vector current. Its connection with the density dependent
chiral (quark) condensate $\langle\bar{\psi}\psi\rangle_\rho$ is determined by the in-medium analogue of the Gell-Mann, Oakes, Renner relation,
\begin{equation}
{f_\pi^*}^2{m_\pi^*}^2 =  -m_q\,\langle\bar{\psi}\psi\rangle_\rho~,
\label{GOR}
\end{equation}
to leading order in the quark mass. Here $m_\pi^*(\rho)$ is the (charge averaged) pion mass
in the medium. A low-density theorem gives the leading $\rho$ dependence of the quark condensate
as
\begin{equation}
\langle\bar{\psi}\psi\rangle_\rho = \langle\bar{\psi}\psi\rangle_0\left(1 - {\sigma_N\over f_\pi^2 m_\pi^2}\,\rho\right)~,
\label{}
\end{equation}
where $\sigma_N = 45\pm8$ MeV is the sigma term of the nucleon.
Assuming that the pion mass is protected by its Goldstone boson nature at
low density, we expect to leading order in the baryon density:
\begin{equation}
{f_\pi^*(\rho)\over f_\pi} \simeq 1 - {\sigma_N\over 2m_\pi^2 f_\pi^2}\,\rho \simeq 1-{\rho\over6\rho_0} \simeq 0.83~
\label{fpi}
\end{equation}
at $\rho = \rho_0 = 0.16$ fm$^{-3}$ and taking $\sigma_N = 45$ MeV for orientation.

A chiral perturbation theory treatment of
in-medium pion dynamics \cite{MOW02}
suggested instead a difference between $m_\pi^*$ and the vacuum pion mass $m_\pi$,
which translates into a stronger density dependence of the pion decay constant, $f_t(\rho)/f_\pi = 1 - (0.26\pm 0.04)\rho/\rho_0$. On the other hand, the charge averaged in-medium pion mass to leading order in the baryon density is given by
\begin{equation}
{m_\pi^*}^2(\rho) = m_\pi^2 -T^{(+)}\,\rho~,
\end{equation}
with the isospin-even forward pion-nucleon amplitude $T^{(+)} = 4\pi(1+m_\pi/M_N)\,a^{(+)}$ taken at threshold, $\omega = m_\pi$. Empirically \cite{Sch99}, the corresponding scattering length $a^{(+)} = (1.6\pm 1.3)\cdot 10^{-3}\,m_\pi^{-1}$ is compatible with zero. This feature derives from a subtle cancellation of non-leading terms which cannot be handled accurately in baryon chiral perturbation theory. Taken as an empirical constraint,  $T^{(+)}(m_\pi)\simeq 0$ implies $m_\pi^*(\rho) \simeq m_\pi$ at low density and hence an approximate scaling of $f_\pi^*$ with the square root of the in-medium chiral condensate as in Eq.(\ref{fpi}). This behavior is actually consistent with the observed energy shifts in deeply bound states of pionic atoms \cite{Su04} and related theoretical calculations \cite{KKW03} (see also Ref.\cite{FG07}).

A recent theoretical study \cite{JHK07} gives further support to these considerations, through a more general derivation of $f_t(\rho)$ which does not have to rely on a detailed evaluation of the charge averaged in-medium pion mass. The basic result of Ref.\cite{JHK07} is
\begin{equation}
f_\pi^*(\rho)\equiv f_t(\rho) = f_\pi \sqrt{{Z\over Z^*}}\,{\langle\bar{\psi}\psi\rangle_\rho \over \langle\bar{\psi}\psi\rangle_0}~,
\end{equation}
where $Z$ and $Z^*$ are the wave function renormalization factors of the pion in vacuum and in-medium, respectively. Their ratio is determined by the pion self-energy $\Pi(\omega, \vec{q},\rho)$,
as follows:
\begin{equation}
{Z\over Z^*} = 1 - {\partial\over\partial\omega^2} \Pi(\omega,\vec{q}=0,\rho)\Big|_{\omega = 0}~.
\end{equation}
Using the low-density expression $\Pi = -T^{(+)}\,\rho$ and the parametrization $T^{(+)}(\omega) = -\sigma_N/f_\pi^2+ \beta\omega^2 + \dots$ one arrives at
\begin{equation}
{f_\pi^*(\rho)\over f_\pi} \simeq 1 - \left({\sigma_N\over m_\pi^2 f_\pi^2}- {\beta\over 2}\right)\rho~,
\end{equation}
to leading order in the density. With the slope $\beta$ determined by the constraint $T^{(+)}(\omega=m_\pi) = 0$ and assuming higher order terms in the expansion of $T^{(+)}$ to be small,  we arrive back at Eq.(\ref{fpi}): $f_\pi^*(\rho_0)/f_\pi = 0.83\pm 0.03$ when the admittedly large uncertainty of the nucleon sigma term is included.

Higher order corrections in the density $\rho$, calculated using in-medium chiral perturbation theory
\cite{KHW08}, can be expressed in terms of a density dependent effective nucleon sigma term with a reduced value at normal nuclear matter density, $\sigma_N^{eff}(\rho_0) = (36\pm 9)$ MeV, leading to a 3-4\% increase of the ratio $f_\pi^*(\rho_0)/f_\pi$ over the value (\ref{fpi}).

Notably, the in-medium QCD sum rule analysis assuming $\sqrt{s_0^*} = 4\pi f_\pi^*$ exhibits chiral
scaling of this sort for the KKW spectral distribution, whereas this is not observed for the RW spectral function.

\section{Note on four-quark condensates}

Given spectral functions which consistently satisfy the sum rules for the zeroth and first moments,
Eqs.(\ref{0momsr},\ref{1momsr}), one can turn to the second moment (\ref{2momsr}) and try to deduce constraints for the four-quark condensate term $c_3$, both in vacuum and in-medium. In particular,
one can discuss deviations from the frequently used factorization assumption for those condensates.
As mentioned, factorization means that the intermediate states produced by the quark operators
entering Eq.(\ref{c_3}) are truncated by the ground state (vacuum) only. Exact factorization means
$\kappa = 1$ in Eqs.(\ref{4q}) and (\ref{c_3fac}).

When performing the consistency analysis including the sum rule (\ref{2momsr}) for the
second moment, it turns
out in all cases that the correction $c_3$ is required to be much larger than the value for a factorized four-quark condensate (with $\kappa = 1$): factorization proves to be unrealistic under any circumstances. For detailed estimates we take a value $\bra\bar{q}q\ket \simeq -(0.2\,\text{GeV})^3$
and find the following results:

i) In vacuum, a lower limit $\kappa \gtrsim 4.5$ is observed which implies strong deviations from factorization.

ii) For both types of spectral functions (KKW and RW) the minimal $\kappa$ required in-medium
(typically $\kappa \gtrsim 3$) is somewhat smaller than in vacuum.

The range of uncertainty is generally large in all cases, with $\kappa$ typically extending from its lower limit up to about twice that value.

One concludes that the four-quark condensates, entering the sum rule
at the level of the 2nd  moment of the spectral function, remain
basically undetermined. This appears to be at variance with reported
attempts to constrain such dimension-six condensates from Borel sum
rules for the nucleon \cite{THK07}. Our findings confirm that the
assumption of ground state saturation for four quark condensates
should be handled  with caution. In the present work the sum
rules are released from such a dispute by restricting procedures to
the 0th and 1st moments of the spectral distribution for which
quantitative statements can indeed be made.

\section{Summary and concluding remarks}

The present work re-emphasizes the usefulness of QCD sum rules for {\cal moments} of spectral
functions (or equivalently, finite energy sum rules), with focus on the $\rho$ meson both in vacuum and in the nuclear medium. The sum rules for the two lowest spectral moments involve only the leading (dimension-four) QCD vacuum condensates as (small) corrections. With inclusion of perturbative QCD terms up to order $\alpha_s^3$, these
sum rules permit an accurate quantitative analysis, unaffected by the large uncertainties from condensates of higher dimension (such as the four-quark condensates).

An important scale parameter in this analysis is the gap
separating low-energy (resonance) and high-energy (continuum)
regions of the spectral function. For the vector-isovector current
correlation function, identifying this gap with the scale for
spontaneous chiral symmetry breaking in vacuum, $4\pi f_\pi$,
reproduces time-honoured current algebra relations and
chiral sum rules characteristic of low-energy QCD. The corresponding
in-medium sum rules for the lowest two spectral moments permit to
address the ``mass shift'' versus ``collisional broadening'' issue
from a new, more quantitative perspective, meaningful even for broad
spectral distributions such as that of the $\rho$ meson at nuclear
matter density. Systematic tests have been performed to confirm
that the conclusions drawn from such analysis do not depend on the detailed
threshold modeling of the transition between resonance and continuum parts
of the spectral distributions, even with strong in-medium broadening.

Two prototype examples of in-medium rho meson spectral functions have been examined from this point of view in the present paper. Both of these show substantial broadening and redistribution of strength into the low-mass region, as compared to the vacuum spectrum. The sum rule analysis of the lowest spectral moments reveals qualitative differences with respect to their Brown-Rho (BR) scaling properties. At the same time, both of these spectral distributions account quite well for the low-mass enhancements observed in dilepton spectra from high-energy nuclear collisions. So one must draw the conclusion that BR scaling can presumably not be tested in such measurements.

Given the consistency constraints derived from the first two sum
rules for the spectral moments, one can then proceed to the third
sum rule equation in this hierarchy (involving the second spectral
moment and QCD condensates of dimension six) and discuss limits for
the four-quark condensates. The outcome of this study demonstrates
that the frequently used factorization approximation for these
condensates is questionable under any circumstances, both in vacuum
and in-medium.

In summary, we repeat that QCD sum rules for the first two moments of vector spectral functions,
when combined with the spontaneous chiral symmetry breaking scale of low-energy QCD, permit
a quantitatively accurate analysis in vacuum, consistent with well established current algebra relations.
The in-medium analogues of these sum rules can be used routinely to clarify and classify the properties of vector meson spectral functions in nuclear matter. An extension to temperature dependent sum rules is in progress with special emphasis on the interesting issue of $\rho - a_1$ mixing in a thermal pionic heat bath.

%%%%%%%%%%%%%%%%%%%%%%%%%%%%%%%%%%
\section*{Acknowledgements}
%%%%%%%%%%%%%%%%%%%%%%%%%%%%%%%%%%
We thank S. Leupold for pointing out a mistake in Ref.\cite{KW99},
which has been corrected in the present paper. One of us (W.W.)
thanks the Yukawa Institute of Theoretical Physics in Kyoto, where
this paper has been finalized, for kind hospitality. He gratefully
acknowledges stimulating discussions with G.E. Brown, T. Hatsuda, V.
Koch and E. Shuryak. We thank R. Rapp for providing his calculated
in-medium spectral distributions. Y.K. is grateful to S.H. Lee and
S.i. Nam for useful discussions.

This work has been supported in part by BMBF, GSI and by the DFG
cluster of excellence Origin and Structure of the Universe. M.P.
acknowledges support by a Feodor Lynen Fellowship from the Alexander
von Humboldt foundation and by U.S. DOE under grant
DE-FG02-94ER40818. M.P. thanks the MIT Center for Theoretical
Physics for hospitality and support.

\begin{appendix}

\section{QCD corrections}

Following Ref.\cite{MW00}, the expression for the $n$-th moment (with $n = 0,1,2$) of the spectral
distribution in the isovector ($\rho$ meson) channel is written
\begin{equation}
\begin{split}
 \int^{s_0}_0ds\,s^n R_\rho(s)&= {s_0^{n+1}\over n+1}\left(c_0+{3\over 2}\varepsilon_n\right)\\
                              &\quad+ (-1)^nc_{n+1}-12\pi^2\,\Pi(0)\,\delta_{n0}~.
 \label{moments}
\end{split}
\end{equation}
The leading perturbative QCD term on the r.h.s. has $c_0 =
{3\over2}\left( 1 + {\alpha_s\over\pi}\right)$. The corrections to
${\cal O}(\alpha_s^3)$ are
\begin{equation}
\varepsilon_n = a_n^{(2)}\left({\alpha_s\over\pi}\right)^2 + \,a_n^{(3)}\left({\alpha_s\over\pi}\right)^3\, ,
\end{equation}
with
\begin{equation}
\begin{split}
 a_n^{(2)}&=1.641 + {2.250\over n+1}~,\\
 a_n^{(3)}&=-10.28 +{11.38\over n+1} + 1.69\left({6\over(n+1)^2}-\pi^2\right)~.
\end{split}
\end{equation}
In applications using (\ref{moments}) the relevant coupling is $\alpha_s(s_0)$ with $s_0\sim 1$
GeV$^2$. In practice we use $\alpha_s(1\,\text{GeV}^2) = 0.50\pm 0.03$ \cite{Be07,PRS07}.

\section{First moment of quark distribution}

An accurate value of $A_1$,
\begin{equation}
 A_1=2\int^1_0 d{x}\,x\big(u+\bar{u}+d+\bar{d}\,\big)~,
\end{equation}
which determines the dominant part of the in-medium modifications in
our sum rule analysis, is obtained from the MRST2001 fits
\cite{MRS01}. In this analysis parton distributions of the proton
are derived from measurements of structure functions by the H1 and
ZEUS collaborations at HERA, and by the D0 and CDF collaborations at
the Tevatron, performing DGLAP evolution. The parametrization of the
parton distributions at $Q^2=1\,\mathrm{GeV}^2$ is:
{\setlength\arraycolsep{2pt}
 \begin{eqnarray}
   x\,u_v&=&0.158\,x^{0.25}(1-x)^{3.33}(1+5.61x^{0.5}+55.49x)~,\nn\\
   x\,d_v&=&0.040\,x^{0.27}(1-x)^{3.88}(1+52.73x^{0.5}+30.65x)~,\nn\\
   xS&=&0.222\,x^{-0.26}(1-x)^{7.10}(1+3.42x^{0.5}+10.30x)~,\nn\\
   x\Delta&\equiv&{x}(\bar{d}-\bar{u})\nn\\
          &=&1.195\,x^{1.24}(1-x)^{9.10}(1+14.05x-45.52x^2)~,\nn\\
   2\bar{u}&=&0.4S-\Delta~,\nn\\
   2\bar{d}&=&0.4S+\Delta~,
  \label{eq:partons}
\end{eqnarray}}
where $u_v$ and $d_v$ denote the valence $u$- and $d$-quark
distributions while $2\bar{u}$ and $2\bar{d}$ are the sea quark
distributions. $\Delta$ denotes the difference between $\bar{d}$ and
$\bar{u}$.

Using this parametrization, $A_1$ at a 1 GeV scale is directly
calculated as
 \begin{equation}
   A_1=2\int^1_0 d{x}\,x\big(u_v+d_v+2\bar{u}+2\bar{d}\,\big)=1.2373~~.
   \label{eq:A1}
 \end{equation}

\end{appendix}

%\newpage

%%%%%%%%%%%%% References %%%%%%%%%%%%%%%%%%%%%%%

\end{document}